\documentstyle[prl,epsfig,aps]{revtex} 
\begin{document} 
 
 
\title{Heated granular fluids: the random restitution coefficient approach} 
 
\author{Alain Barrat$^1$, Emmanuel Trizac$^1$ and Jean-No\"el Fuchs$^2$} 
 
\address{  
$^1$ Laboratoire de Physique Th{\'e}orique 
\cite{umr}, B{\^a}timent 210, Universit{\'e} 
de Paris-Sud, 91405 Orsay, France \\ 
$^2$ Laboratoire Kastler-Brossel\cite{umr2}, 
D\'epartement de Physique de l'E.N.S, 24 rue Lhomond, 75231 Paris, France \\
} 
 
\date{\today} 
 
\maketitle 
\begin{abstract} 
We introduce the model of inelastic hard spheres with random restitution 
coefficient $\alpha$, in order to account for the fact that, in a vertically
shaken granular system interacting elastically with the vibrating boundary, 
the energy injected vertically is transferred
to the horizontal degrees of freedom through collisions only, 
which leads to heating through collisions,
i.e. to inelastic horizontal collisions with an effective restitution coefficient 
that can be larger than $1$. This allows the system to 
reach a non-equilibrium steady state, where we
focus in particular on the single particle velocity distribution $f(v)$
in the horizontal plane, and
on its deviation from a Maxwellian. Molecular Dynamics simulations
and Direct Simulation Monte Carlo (DSMC) show that,
depending on the distribution of $\alpha$, different shapes of $f(v)$ can be
obtained, with very different high energy tails.
Moreover, the fourth cumulant of the
velocity distribution quantifying the deviations from Gaussian
statistics is obtained analytically from the Boltzmann equation and 
successfully tested against the simulations.
\end{abstract} 

\vskip .5pc 

\section{Introduction}

Despite a wealth of recent experimental investigations, it seems that 
there is no consensus
yet as to the characteristics of the collective behaviour of vibrated
granular mono- or multi-layers \cite{Clement,Warr,Olafsen,Losert,Rouyer,Kudrolli}. 
Among the statistical properties of interest,
the velocity fluctuations and more precisely the deviations from 
Maxwell-Boltzmann distribution due to inelastic collisions
have been intensively studied for rapid granular flows, and several groups
reported an overpopulated high energy tail that can be fitted by stretched
exponentials with various exponents. 

Theoretical investigations commonly use the
inelastic hard spheres (IHS) model \cite{Luding,TwanPre}, 
which has proven 
very useful despite the simplicity of its 
definition: smooth hard spheres undergo binary inelastic 
momentum conserving collisions,
thereby losing during the collision a constant fraction of their 
relative normal velocity (and therefore loosing energy).
Many studies have concentrated 
on the homogeneous cooling state of the IHS
\cite{goldhirsch,mcnamara,brey1,esipov,brilliantov,huthmann-orza-brito},
obtained by letting the system
evolve without any energy injection.
On the other hand, experiments on strongly vibrated granular media 
consider systems that 
are heated by an external forcing (e.g. a vertically oscillating plate)
to compensate for the energy loss due
to collisions, and are therefore in a non equilibrium stationary
state (NESS). 
Various ways of modeling the heating mechanism have been put forward, mostly
consisting in the study of an effective system with a constant coefficient of
normal restitution $\alpha$, in which
the energy lost through collisions is balanced by energy 
injected by random ``kicks'' \cite{Williams,Puglisi,twan,Pre,Cafiero}. 
In this context, analytical and numerical
results have been obtained for the high energy tail of the velocity
distributions and for its fourth cumulant, a natural
measure of the deviation from a Gaussian distribution
\cite{esipov,twan,montanero}. Moreover, more realistic Molecular Dynamics 
simulations of vibrated soft sphere mono-layers \cite{Nie} 
have been able to reproduce the
phenomenology obtained in some experiments \cite{Olafsen}.

In this paper, we propose 
an alternative modeling of a three-dimensional system shaken
vertically along the $z$ axis, for which the velocities are studied 
both numerically and analytically in the
horizontal ($xy$) plane. 
If the collisions with the boundaries are elastic, 
the vibrating wall feeds energy in the system in the $z$ direction 
only, but not in the $xy$ plane where the correponding
two dimensional energy is either lost or gained at each 
collision between two grains. Upon restricting to the $xy$ velocities,
we obtain a system that is subject to an effective stationary dynamics 
with sequential energy injection or dissipation consecutive to
particle-particle
collisions only. We shall thus disregard the vibrating boundary
and concentrate on the effective planar dynamics where 
energy gains and losses statistically cancel in the NESS,
so that an effective restitution coefficient 
can either be larger or smaller than $1$;
this will be accounted for by a {\it random} $\alpha$, drawn from
a probability distribution $\rho(\alpha)$
(such that the second moment $\overline{\alpha^2}=1$ in order to conserve
energy on average). Our approach, in which momentum transfer occurs
only through collisions, consequently 
differs from the situations investigated
in \cite{Williams,Puglisi,twan,Pre,Cafiero} where energy is injected
globally into the system at regular time intervals through a stochastic
external force.
Clearly, the functional form of $\rho(\alpha)$
reflects the energy injection mechanism, and it is a difficult
task to establish this connection. The physical situations
that could be described by the present model are mono- or multi-layers
of grains provided the interactions with the wall are elastic, or
a multilayer system with arbitrary interactions with the wall,
in a ``bulk'' region far from the boundaries where energy and density
can be considered as constant.
In a very different context, a similar stochastic 
coefficient of restitution has been introduced to study the dynamics of
a one-dimensional granular gas, thereby accounting for 
internal degrees of freedom \cite{Aspelmeir}.

The paper is organized as follows: after setting the general
framework in section II, we consider the 
two-dimensional projection of three-dimensional simulations
of inelastic hard spheres with constant restitution
coefficient $\alpha$, energy being injected in the third direction
(section III). We 
analyze in particular the distribution of
energy transfer through collisions. 
In section IV, the two dimensional IHS model with random 
restitution coefficient is then
studied by analytical and numerical means.
Section V is devoted to a short investigation of the one-dimensional
case, and some conclusions are finally presented in section VI.

\section{General framework}

\subsection{IHS: definitions and notations}

In the inelastic hard spheres (IHS) model, 
grains are modeled as smooth hard spheres of mass $m$ undergoing
binary, inelastic and momentum-conserving collisions: a collision
between two spheres labeled by $1$ and $2$, with velocities 
${\mathbf v}_1$ and ${\mathbf v}_2$,
dissipates a fraction $(1-\alpha)$ of the component of
the relative velocity ${\mathbf v}_{12}={\mathbf v}_1 -{\mathbf v}_2$ 
along the center-to-center direction $\bbox{\hat \sigma}$. 
Noting with stars the post-collision velocities, this translates
into ${\mathbf v}_{12}^* \cdot \bbox{\hat{\sigma}}
= -\alpha \,{\mathbf v}_{12} \cdot \bbox{\hat{\sigma}}$, while 
the tangential relative velocity (perpendicular to $\bbox{\hat\sigma}$) 
is conserved, i.e.:
\begin{eqnarray}
{\mathbf v}_1^{*}&=& {\mathbf v}_1 - \frac{1}{2}\left(1+\alpha\right)
({\mathbf v}_{12}\cdot \bbox{\hat{\sigma}}) \bbox{\hat{\sigma}}   \nonumber \\
{\mathbf v}_2^{*}&=& {\mathbf v}_2 + \frac{1}{2}\left(1+\alpha\right)
({\mathbf v}_{12}\cdot \bbox{\hat{\sigma}}) \bbox{\hat{\sigma}} \ .
\end{eqnarray}
These equations are also sometimes written in terms of 
restituting collisions, i.e. for collisions which yield
$({\mathbf v}_1,{\mathbf v}_2)$ as postcollisional velocities, for
precollisional velocities $({\mathbf v}_1^{**},{\mathbf v}_2^{**})$:
\begin{eqnarray}
{\mathbf v}_1^{**}&=& {\mathbf v}_1 - 
\frac{1}{2}\left(1+\frac{1}{\alpha}\right)
({\mathbf v}_{12}\cdot \bbox{\hat{\sigma}}) \bbox{\hat{\sigma}} \nonumber \\
{\mathbf v}_2^{**}&=& {\mathbf v}_2 + 
\frac{1}{2}\left(1+\frac{1}{\alpha}\right)
({\mathbf v}_{12}\cdot \bbox{\hat{\sigma}}) \bbox{\hat{\sigma}} \ .
\end{eqnarray}

\subsection{Evolution equation}

The Enskog-Boltzmann equation describes the evolution of the one-particle
distribution function $f({\mathbf r},{\mathbf v},t)$, upon the
molecular chaos hypothesis \cite{Resibois}. 
In the homogeneous case, for hard spheres
of diameter $\sigma$, in $d$ dimensions, this equation reads:
\begin{eqnarray}
\frac{\partial f({\mathbf v}_1,t)}{\partial t} &=&
\chi \sigma^{d-1} \int d{\mathbf v}_2
\int' d\bbox{\hat{\sigma}} ({\mathbf v}_{12} \cdot \bbox{\hat{\sigma}})
\left\{
\frac{1}{\alpha^2}f({\mathbf v}_1^{**},t)f({\mathbf v}_2^{**},t) -
f({\mathbf v}_1,t)f({\mathbf v}_2,t) \right\} \nonumber \\
 &\equiv& \chi I(f,f)
\label{eq:ensk}
\end{eqnarray}
The prime on the integration symbol is a shortcut for
$\int d\bbox{\hat{\sigma}} 
\Theta ({\mathbf v}_{12} \cdot \bbox{\hat{\sigma}})$
where $\Theta$ is the Heavyside function, while
$\chi$  accounts for excluded volume effects (for elastic hard spheres,
$\chi$ coincides with the density dependent pair correlation 
function at contact). 
The forcing mechanism necessary to sustain a NESS can be of 
different types, and has not been introduced in Eq. (\ref{eq:ensk}). 
This issue will be addressed in section IV.

\subsection{Sonine expansion for the velocity distribution}

Because of analytical, numerical and experimental evidences, it is 
customary to look for scaling solutions of eq. (\ref{eq:ensk}), in the form
\cite{goldshtein}
\begin{equation}
f({\mathbf v},t) = \frac{n}{v_0^d(t)} 
\tilde{f}\left(\frac{v}{v_0(t)}\right) \ ,
\end{equation}
where $n$ is the density and
the thermal velocity $v_0$ is by definition related to the temperature $T(t)$
through
$\frac{m}{2}v_0^2(t)=T(t)$, where in turn the temperature is defined by the
average kinetic energy of the particles:
\begin{equation}
\frac{d n}{2}T(t)=
\int d{\mathbf v} \frac{m}{2}v^2 f({\mathbf v},t) \ .
\end{equation}

Replacing the scaling function in eq (\ref{eq:ensk}),
together with the law of evolution of the temperature 
(see \cite{twan} for details) yields the following equation:
\begin{equation}
\frac{\mu_2}{d} \left( d + c_1 \frac{d}{dc_1} \right) \tilde{f}(c_1)
= \tilde{I}(\tilde{f},\tilde{f}) \ ,
\label{eq:ftilde}
\end{equation}
where ${\mathbf c}_i = {\mathbf v}_i/v_0(t)$, 
\begin{equation}
\tilde{I}(\tilde{f},\tilde{f}) = 
\int d{\mathbf c}_2
\int' d\bbox{\hat{\sigma}} ({\mathbf c}_{12} \cdot \bbox{\hat{\sigma}})
\left\{
\frac{1}{\alpha^2}\tilde{f}({c}_1^{**})\tilde{f}({c}_2^{**}) 
- \tilde{f}({c}_1)\tilde{f}({c}_2) \right\} \ ,
\label{eq:itilde}
\end{equation}
and 
$\mu_p \equiv - \int d{\mathbf c}_1 c_1^p \tilde{I}(\tilde{f},\tilde{f})$.

In the elastic case ($\alpha=1$), $\tilde{f}$ is the
Gaussian $\Phi(c)=\pi^{-d/2} \exp (-c^2)$ (since $\tilde{I}(\Phi,\Phi)=0$).
The deviations from $\Phi$ are
studied by an expansion in terms of Sonine polynomials \cite{Landau}
\begin{equation}
\tilde{f}(c) = \Phi(c) \left( 1 + \sum_{p=1}^{\infty} a_p S_p (c^2)\right) \ .
\label{eq:sonine}
\end{equation}
Indeed, the coefficients $a_p$ can be obtained from the moments of $\tilde{f}$,
because the $S_p$ satisfy the orthogonality relations:
\begin{equation}
\int d{\mathbf c} \Phi(c) S_p(c^2) S_{p'}(c^2) = \delta_{pp'} N_p
\end{equation}
where the $N_p$ are normalization constants. These polynomials consequently depend
on the dimension $d$; the first ones read:
\begin{eqnarray}
S_0(x)&=&1 \nonumber \\
S_1(x)&=&-x + \frac{d}{2}  \nonumber \\
S_2(x)&=& \frac{x^2}{2} - \frac{d+2}{2} x + \frac{d(d+2)}{8}
\end{eqnarray}
The orthogonality properties allow to write
\begin{equation}
a_p = \frac{1}{N_p} \langle S_p (c^2) \rangle .
\end{equation}
In particular, $a_1= (2/d)(-\langle c^2 \rangle + d/2)$ vanishes because
of the definition of temperature, and $a_2$ is related to the fourth cumulant
of $\tilde{f}$:
\begin{equation}
a_2 = \frac{4\langle c^4 \rangle }{d(d+2)} -1 \ .
\end{equation}
In section IV, we shall briefly recall the different steps 
involved in the computation of $a_2$ that quantifies the deviations
from gaussianity.

\subsection{DSMC Method}

The Direct Simulation Monte Carlo (DSMC) method allows 
to solve numerically the Boltzmann equation (\ref{eq:ensk}) \cite{Bird}.
It has been successfully used e.g. for the study of the 
homogeneous cooling state of the IHS, 
in \cite{brey1} to assess
the validity of the Sonine expansion, and
in \cite{brey2} to analyze the high energy tail of $\tilde{f}$. In 
\cite{montanero}, the case of heated granular fluids has also been 
considered and compared to the theoretical predictions of \cite{twan}.

The positions of the particles
do not appear in equation (\ref{eq:ensk}). Therefore all
reference to space is here useless; this amounts, in the
usual DSMC language, to taking only one cell
where all particles stand, and of course eliminates the
possibility to study spatial inhomogeneities. Moreover, upon an 
appropriate rescaling of time, we shall take 
$\chi=1$ and $\sigma=1$.
Since no external force is present, the velocities of the particles
do not change between collisions, and no ``free streaming stage'' is here
necessary, and the simulation has the following scheme:
we start from $N$ particles with random velocities taken from
an arbitrary distribution (e.g. Gaussian, or flat); then
the evolution proceeds by choosing at random
pairs of particles $(i,j)$, a direction $\bbox{\hat{\sigma}}$ for
their center to center direction,
and updating their velocities according to the
collision rule with a probability proportional to
$\Theta({\mathbf v}_{ij} \cdot \bbox{\hat{\sigma}})
{\mathbf v}_{ij} \cdot \bbox{\hat{\sigma}}$. Once a stationary state
is reached, running averages can be taken on the quantities of interest.

\subsection{Molecular Dynamics simulations}

The Molecular Dynamics (MD) simulations 
integrate the exact equation of motion of the model,
with no reference to the Boltzmann equation: we consider
$N$ spheres of diameter $\sigma$, in a box of linear size $L$
in dimension $d$, with periodic boundary conditions \cite{Herrmann,Allen}. 
These spheres initially 
have  random velocities, and we use an 
event-driven algorithm to study their dynamics. Once more, running
averages are taken once a stationary state is reached.

\section{2D Projection of a 3D system}

We consider an IHS model with constant (velocity independent)
restitution coefficient $\alpha<1$,
in dimension $d=3$, the Cartesian coordinates being
labeled $x,y$ and $z$. The collisions are dissipative
but energy is re-injected in the vertical direction ($z$) in the
following way:
after each collision, the $z$-component of the velocity
of the particles having collided is randomly drawn from
a Gaussian distribution at fixed temperature, while the $x-y$ 
components remain those resulting from the initial inelastic collision.
This procedure is intended to mimic both the energy injection
due to a vibrating wall, and transfer to horizontal
degrees of freedom through collisions only: indeed, a vibrating boundary
can yield in the bulk of a multi-layer system an equilibrated Gaussian
vertical velocity~\cite{Warr,Helal}. We could have used
different vertical velocity distributions, in particular asymetric ones,
with the same conclusion: the point here
is to illustrate the horizontal energy tranfer mechanism.

\begin{figure}
\centerline{
       \psfig{figure=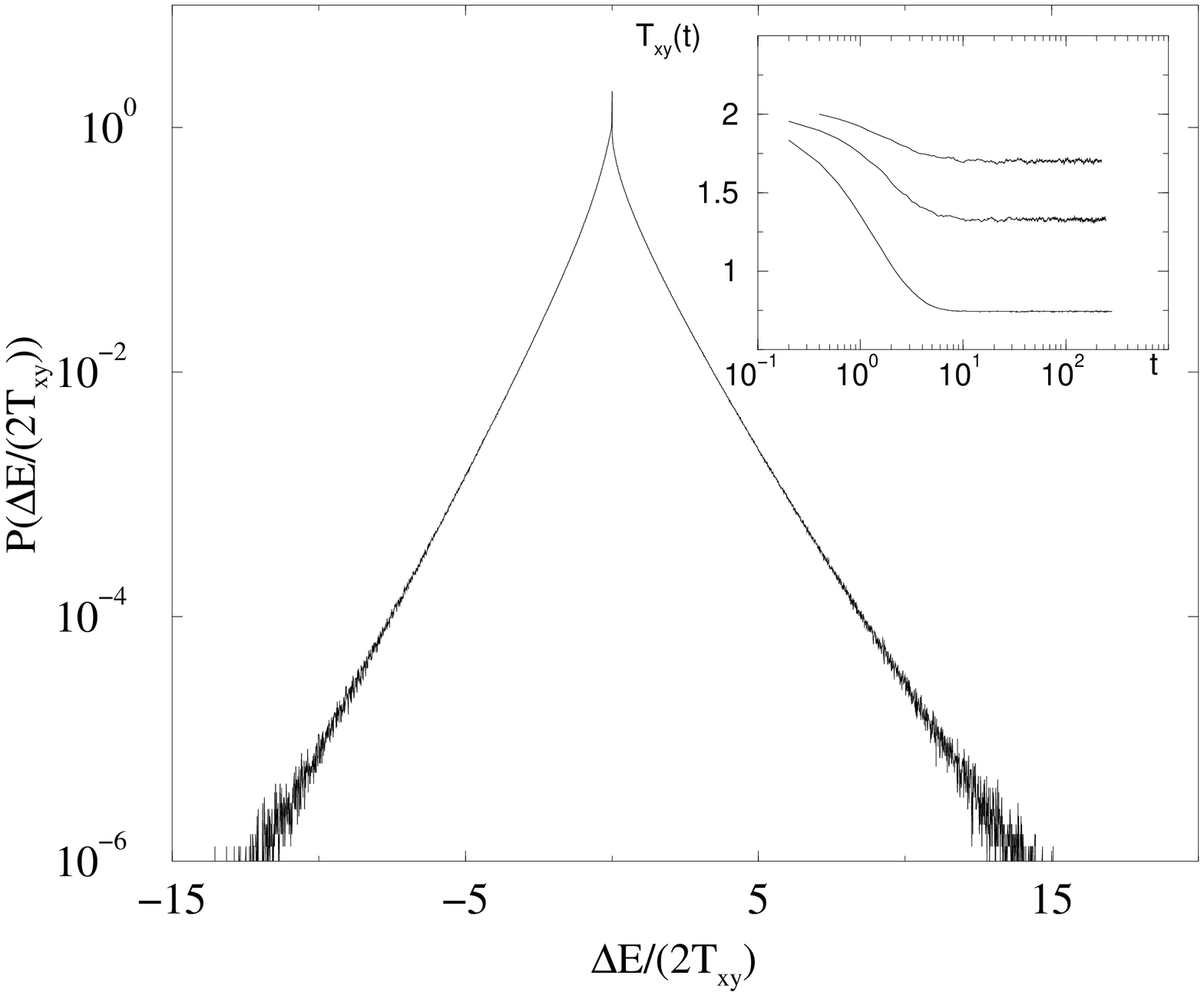,width=6cm,angle=-90}
  \vspace{0.5cm}}
\caption{Histogram of the two-dimensional energy change consecutive to a 
collision for a 3-D DSMC simulation of a system of
$N= 5\,10^5$ particles, with constant
restitution coefficient $\alpha=0.9$ and energy injection in the $z$-direction.
The positive part corresponds to effective restitution
coefficients larger than $1$.
The inset shows the transient approach to the NESS of the
two-dimensional temperature
for $\alpha=1,\ 0.9,\ 0.6$ (from top to bottom). The initial value of the 
temperature coincides with that of the equilibrated vertical degrees
of freedom. A similar anisotropy 
between horizontal and vertical temperatures has been observed in
\protect\cite{Nie}
}
\label{fig:deltaeproj}
\end{figure}

Let us indeed analyze the system in the horizontal plane ($xy$).
After a transient, the two-dimensional temperature 
$T_{xy}=\frac{1}{2}(\langle v_x^2 + v_y^2\rangle)$ remains stationary
(see the inset of Fig. \ref{fig:deltaeproj}): 
overall, i.e. when the three components of the velocities are considered,
a collision is dissipative, but in the $xy$-plane, energy can 
be gained by a transfer from the $z$-direction. In the horizontal plane, 
both phenomena compensate
each other, as appears in Fig. \ref{fig:deltaeproj}: the histogram of the
energy transferred in the $xy$-plane at each collision
(given by $\Delta E = v_{1x}'^2 + v_{1y}'^2 + v_{2x}'^2 + v_{2y}'^2
-( v_{1x}^2 + v_{1y}^2 + v_{2x}^2 + v_{2y}^2)$, where $1$ and $2$ label
the two colliding particles)
has of course a negative part but also a positive one, 
corresponding to an energy gain in the $xy$-plane thought a three-dimensional 
collision (the positive part would be absent in a simulation
without heating and constant $\alpha < 1$). This positive part compensates 
the energy loss and allows to reach a NESS with a constant $T_{xy}$.

It would be feasible to inject energy in a more refined way, by considering for
instance the collisions with the vibrating wall. Seeking for analytical results
concerning the 2D horizontal velocity distribution, we shall however
make the simplifying assumption that the effective 2D restitution coefficient
decouples from the impact relative velocity. The resulting
zeroth order modeling introduced below displays qualitatively
the same energy transfer behaviour as the projected 3D system
and is amenable to a kinetic theory description.

\section{Random restitution coefficient}

We shall hereafter consider an IHS model for which, at each
collision, the restitution coefficient is
drawn from a probability distribution $\rho(\alpha)$. The means
over $\rho(\alpha)$ will be denoted by an overline, and
the distribution of $\alpha^2$ by $\tilde{\rho}(\alpha^2)$.
Since, in a binary collision with restitution coefficient
$\alpha$, the energy change is:
\begin{equation}
\Delta E = \frac{m}{2} ({\mathbf v}_{12}^{**} \cdot  \hat{\bbox \sigma})^2
\frac{(\alpha^2 -1)}{2} \ ,
\end{equation}
(where ${\mathbf v}_{12}^{**}$ is the relative velocity before collision, 
$\hat{\bbox \sigma}$
the center to center direction, and $m$ the mass of the particles),
and since the value of $\alpha$ is taken uncorrelated with the velocities
of the particles, we shall consider distributions 
with $\overline{\alpha^2}=1$ in order to ensure a stationary, constant
temperature regime (at each collision, energy changes, but  
is conserved on average: 
$\overline{\Delta E}=0$). Moreover, we will restrict ourselves
to positive values of $\alpha$. Since the average energy is constant, the
granular temperature is also a constant determined by the initial
velocity distribution. This model is therefore intended to study
the distribution of {\em rescaled} velocities ${\mathbf c=v}/v_0$.

\subsection{Analytical results}

The methods of \cite{twan} for the case
of constant normal restitution (with or without external heating) 
can be easily applied to the case of random $\alpha$
to systematically
obtain the coefficients of the Sonine expansion. We refer to~\cite{twan} for
details and sketch only the principal steps. 
Note that the calculation
is made under the hypothesis of a small $a_2$: 
the Sonine expansion (\ref{eq:sonine}) is therefore truncated after second order
and besides, terms of order $a_2^2$ are discarded (it is possible
to go beyond the linear approximation in $a_2$,
with again truncation of (\ref{eq:sonine}) after $n=2$; it was however shown 
that in the case of constant
$\alpha$, the correction is less than $10\%$~\cite{brilliantov}).

Once the moments 
$\mu_p \equiv - \int d{\mathbf c}_1 c_1^p \tilde{I}(\tilde{f},\tilde{f})$
have been defined, multiplying the equation (\ref{eq:ftilde}) 
by $c_1^p$ and integrating over $c_1$ yields
\begin{equation}
\mu_p = \mu_2 \frac{p}{d} \langle c^p \rangle.
\label{eq:mupmu2}
\end{equation}
Taking $p=4$ and approximating $\tilde{f}$ by its second order Sonine 
expansion, it is now possible to evaluate $\mu_2$, $\mu_4$ and 
$\langle c^4 \rangle$, that can be averaged over $\alpha$:
\begin{eqnarray}
\langle c^4 \rangle &=& \frac{d(d+2)}{4} (1+a_2) \nonumber \\
\overline{\mu_2} &=& \frac{\Omega_d}{2 \sqrt{2\pi}} 
(1-\overline{\alpha^2})\left( 1+\frac{3}{16}a_2 \right) \nonumber \\
\overline{\mu_4} &=& \sqrt{\frac{2}{\pi}} \Omega_d
(\overline{T_1} + a_2 \overline{T_2}) \ ,
\end{eqnarray}
with $T_1 = (1-\alpha^2)(d+3/2+\alpha^2)/4$,
$T_2= 3(1-\alpha^2)(10d + 39 + 10\alpha^2)/128 + (1+\alpha)(d-1)/4$
($\Omega_d$ is the volume of the $d$-dimensional unit sphere).
Inserting these relations in (\ref{eq:mupmu2}) and neglecting terms
of order $a_2^2$ leads to the final expression for
$a_2$ in dimension $d$:
\begin{equation}
a_2 = 16 \frac{1 - 3\overline{\alpha^2} +2 \overline{\alpha^4}}
{9 + 24 d  +32 (d-1) \overline{\alpha} + (8d - 11)\overline{\alpha^2} 
-30 \overline{\alpha^4}} \ .
\label{eq:a2}
\end{equation}
It can be checked that the expression obtained in~\cite{twan}
is recovered in the case of constant $\alpha$: when 
$\rho(\alpha)=\delta(\alpha-\alpha_\star)$, we get 
\begin{equation}
a_2 = 16 \frac{1 - \alpha_\star -2 \alpha_\star^2 +2 \alpha_\star^3}
{9 + 24 d  + 8\alpha_\star d - 41\alpha_\star +30(1-\alpha_\star)\alpha_\star^2} \ .
\end{equation}
Moreover, in the case $\overline{\alpha^2}=1$, expression (\ref{eq:a2}) 
reduces to
\begin{equation}
a_2 = 16 \frac{\overline{\alpha^4} -1}{16 d -1 +16 (d-1) \overline{\alpha} 
-15 \overline{\alpha^4}} \ .
\label{eq:a2bis}
\end{equation}
The values of $a_2$ corresponding to the
various distributions of normal restitutions shown in
Fig. \ref{fig:rhoalpha}, are given in appendix A.

An important consequence of Eq. (\ref{eq:a2}) is that the fourth
cumulant depends
only on $\overline{\alpha}$,
$\overline{\alpha^2}$, and $\overline{\alpha^4}$: two different
distributions $\rho(\alpha)$ having the same first, second and fourth
moments should then yield very similar velocity distributions. This will
be checked numerically in the next subsection and can be considered 
as a test for the consistency of the linear order approximation in $a_2$
underlying the analytical computation.

\begin{figure}
\centerline{\psfig{figure=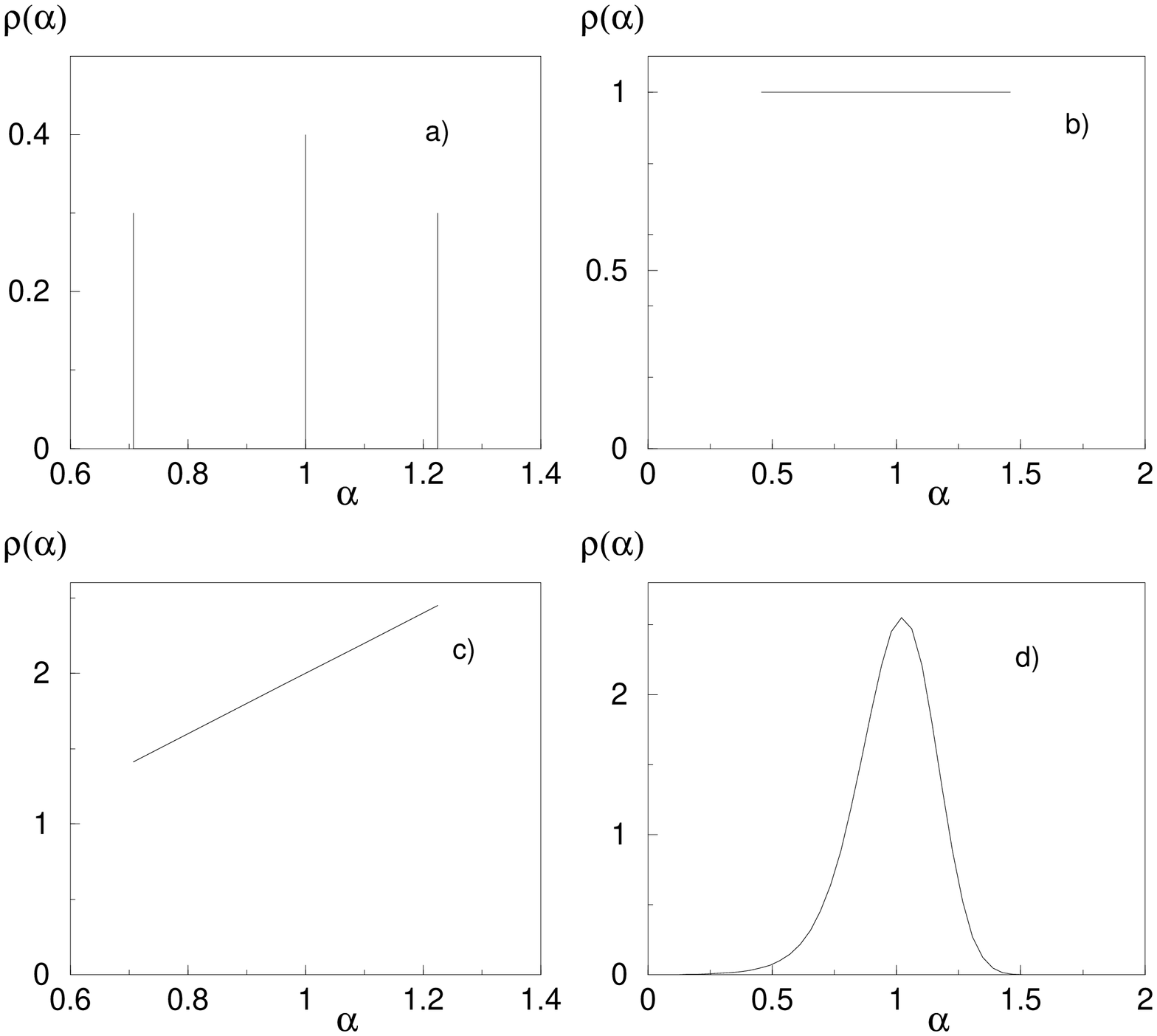,width=6cm,angle=-90}
\vspace{0.5cm}} 
\caption{Various distributions $\rho(\alpha)$
of the restitution parameter, 
fulfilling $\overline{\alpha^2}=1$:\\ 
a) trimodal distribution
 $\rho(\alpha)=\frac{b}{2} 
[\delta(\alpha - \sqrt{1+\gamma}) +
\delta(\alpha - \sqrt{1-\gamma})] + (1-b)\delta(\alpha - 1)$,
with the bimodal case included for $b=1$; b) flat $\rho(\alpha)$
($\alpha \in [0.457427;1.457427]$);
c) flat $\tilde{\rho}(\alpha^2)$ (i.e. linear $\rho(\alpha)$), for
$\alpha^2 \in [1-\gamma,1+\gamma]$;
d) Gaussian $\tilde{\rho}(\alpha^2)$.}
\label{fig:rhoalpha}
\end{figure} 

The question of the high energy tail, which was
addressed in~\cite{esipov,twan} by neglecting the gain term in the
collision integral $\tilde{I}$, turns out to be more problematic. 
Whenever $\alpha$ is allowed to take values exceeding 1, the gain term
can no longer be discarded and we could not obtain analytical predictions.
We shall therefore resort 
to a numerical resolution of the Boltzmann equation and to molecular dynamics
simulations to analyze the high velocity statistics.
It can however be shown that the velocity
distribution is Gaussian in the elastic case only 
[i.e. for $\rho(\alpha)=\delta (\alpha -1)$, see appendix B].

\subsection{Numerics}

We have simulated the IHS model with random restitution coefficient
in two dimensions
for various distributions $\rho(\alpha)$, both with DSMC and MD methods.
The DSMC simulations were performed with $3.10^5$ particles,
the MD ones with up to $50000$ particles, and packing fractions
from $10\%$ to $40\%$. The restitution
coefficient $\alpha$ were drawn from distributions of various types:
bimodal or trimodal distributions, flat distributions,
flat distributions of $\alpha^2$, Gaussian distributions of $\alpha^2$
(with a cutoff in zero to ensure that $\alpha^2 \ge 0$). The distributions 
are shown in Fig. \ref{fig:rhoalpha}: no large values of $\alpha$ or
pathological distributions will be used.

\begin{figure}
\centerline{
       \psfig{figure=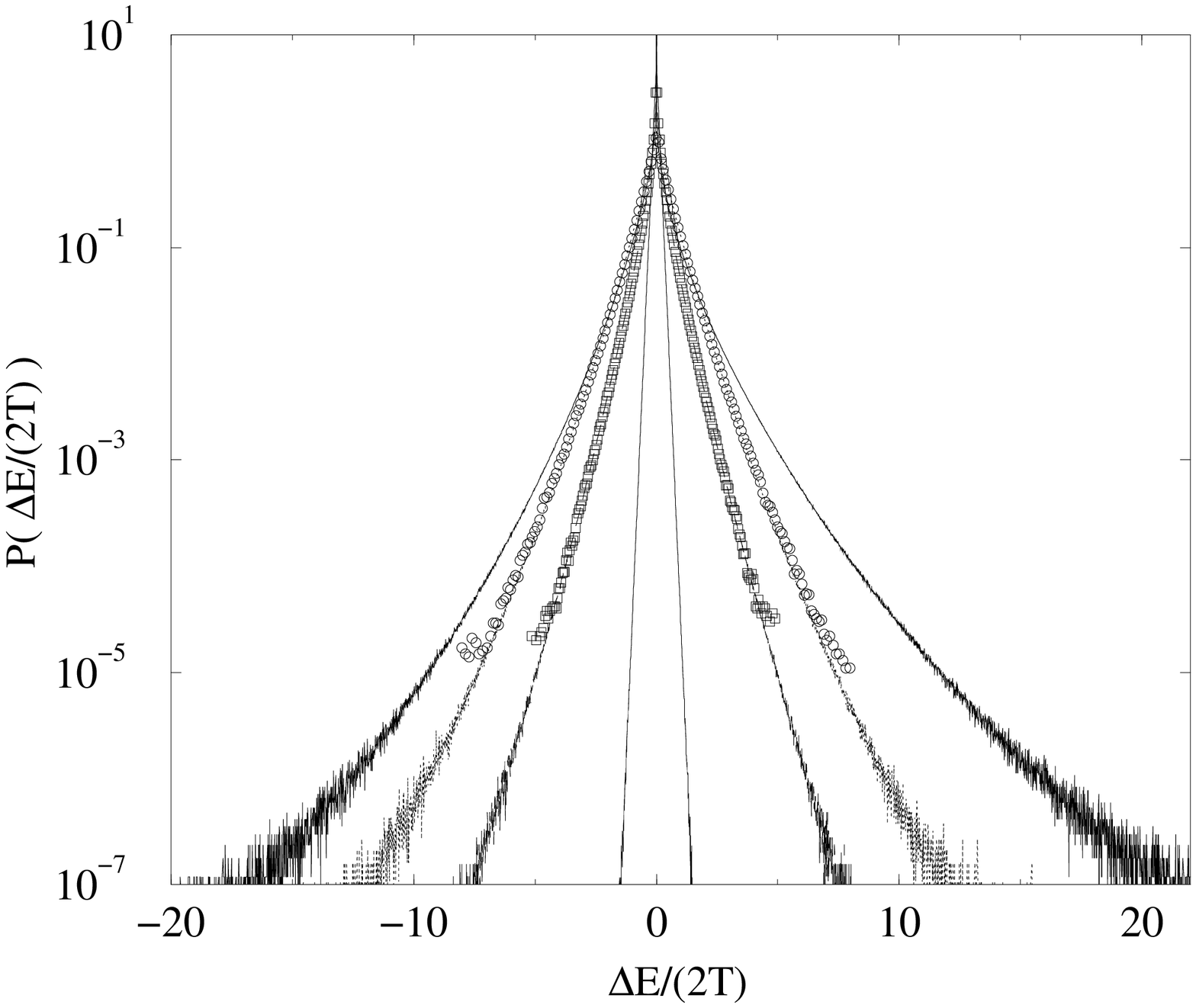,width=6cm,angle=-90}
  \vspace{0.5cm}}
\caption{MD (symbols) and DSMC (lines) simulations of the random $\alpha$
model. MD curves with packing fraction $10\%$.
The various curves correspond to various $\rho(\alpha)$:
from top to bottom (on the right), flat $\rho(\alpha)$ (DSMC), bimodal
$\tilde{\rho}(\alpha^2)= \frac{1}{2}(\delta(\alpha^2 - 1/2)
+\delta(\alpha^2 - 3/2))$ (DSMC and MD), 
uniform $\alpha^2 \in [0.5;1.5]$ (DSMC and MD), and
bimodal $\rho(\alpha)=\frac{1}{2}(\delta(\alpha - 1.04)
+\delta(\alpha - 0.958332)$ (DSMC).
}
\label{fig:deltaemddsmc}
\end{figure}

The stationarity of the kinetic energy (following from 
$\overline{\alpha^2} = 1$)
is controlled during the simulation, and allows to obtain in a single run 
the velocity distribution with 
a precision of typically $7$ to $8$ orders of magnitude.
We first show in Fig. \ref{fig:deltaemddsmc}
the histogram of the energy transfers
during the collisions, corresponding to various $\rho(\alpha)$, for
both  DSMC and MD simulations; the similarity with the results 
of the 3 dimensional simulations (Fig. \ref{fig:deltaeproj}) allows 
to validate the model as far as energy transfer is concerned
(the precise form of the histogram depends on the
probability distributions $\rho(\alpha)$, but the qualitative features
are those displayed by the projected 3D system considered in section III).

\begin{figure}
\centerline{
       \psfig{figure=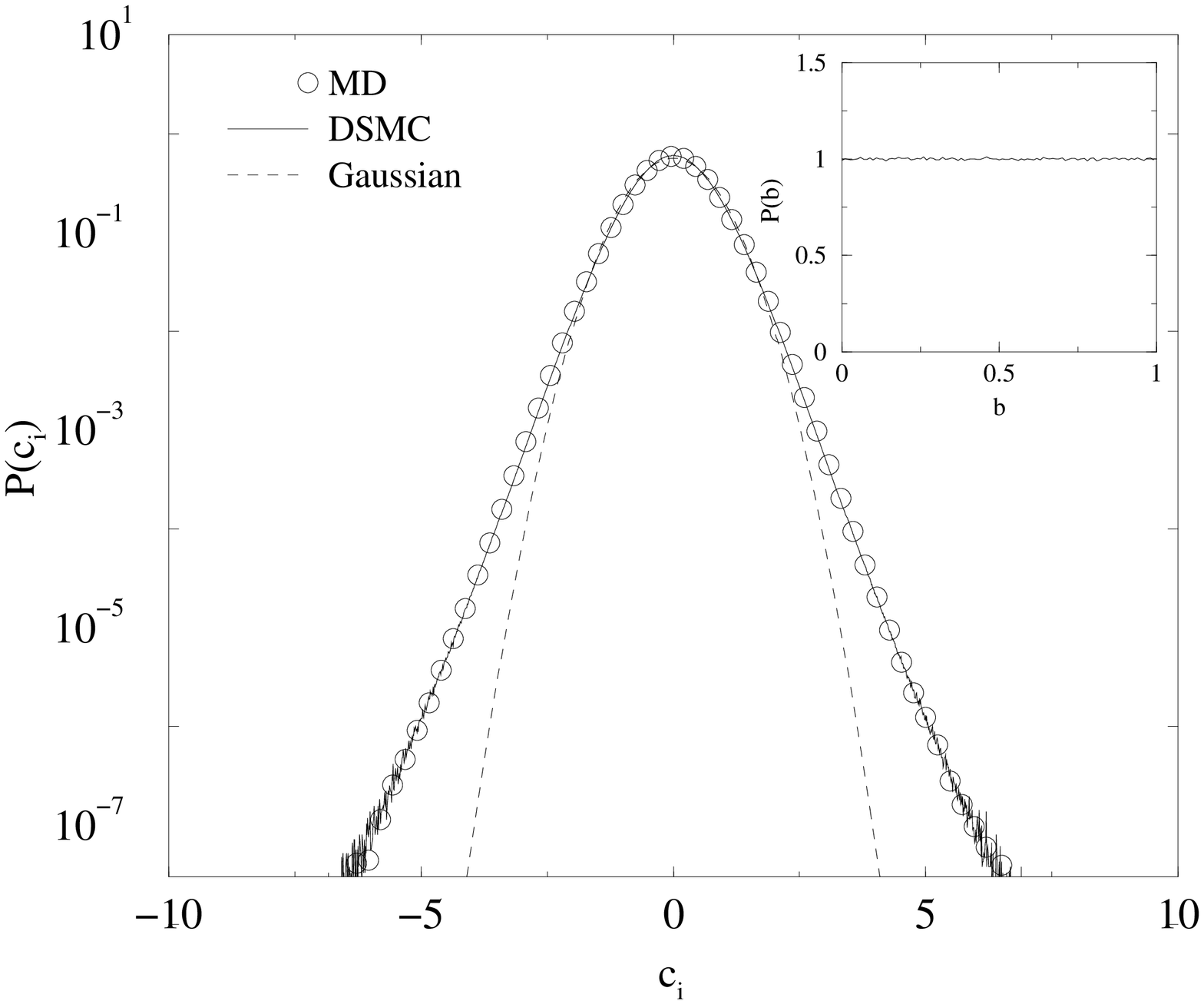,width=6cm,angle=-90}
  \vspace{0.5cm}}
\caption{Velocity distributions for the random $\alpha$ model,
for MD (circles)  and DSMC (solid line) 
simulations with respectively $50000$ and $300000$ particles,
with the same $\rho(\alpha) [$flat $\tilde{\rho}(\alpha^2)$ between 
$0$ and $2$]. The agreement between both sets of data is striking.
Large deviations from the Maxwellian (dotted line) are observed.
The inset is the distribution of impact parameters for the MD simulations,
showing no violation of molecular chaos.
}
\label{fig:compmddsmc}
\end{figure}

Figure \ref{fig:compmddsmc}
shows the velocity distributions
for a flat $\tilde{\rho}(\alpha^2)$ between $0$ and $2$, 
for DSMC and MD simulations. The agreement
between both sets of data is remarkable, and 
was also checked for other choices of $\rho(\alpha)$. 
Moreover, the curves obtained in MD simulations with small or large
packing fractions (up to 40\%) are indistinguishable (not shown).
The inset shows that the distribution of
impact parameters in Molecular Dynamics simulations
is flat, which is a hint that no violation of molecular
chaos is observed and an indication that the factorization 
of the 2-particle correlation function resulting in
Eq. (\ref{eq:ensk}) holds.
The super-imposition of MD (where {\it a priori} 
inhomogeneities and/or violations
of molecular chaos could appear), and DSMC results 
is in contrast with the phenomenology at constant $\alpha$ 
\cite{SLuding} or with randomly driven IHS~\cite{Ignacio}, and is probably
due to an efficient randomization of the velocities with the collision
rule of the present model. For a constant dissipative restitution
parameter, colliding particles emerge with more parallel velocities 
than in the elastic case $\alpha=1$. When they recollide, their velocities
are still more parallel. Here, this mechanism for the creation of
velocity correlations violating molecular chaos
seems removed by the possibility of having $\alpha>1$. This validates the
theoretical approach based on the Boltzmann equation.

The remainder of this article is devoted to the deviations from gaussianity
that can be seen in Fig. \ref{fig:compmddsmc}. To this aim, we shall use DSMC
simulations that allow to obtain precise velocity distributions
for a larger range of velocities than the MD method. However, as stated above, 
we always observed an excellent agreement between DSMC and MD velocity
statistics, up to the resolution of MD. 

\begin{figure}
\centerline{
       \psfig{figure=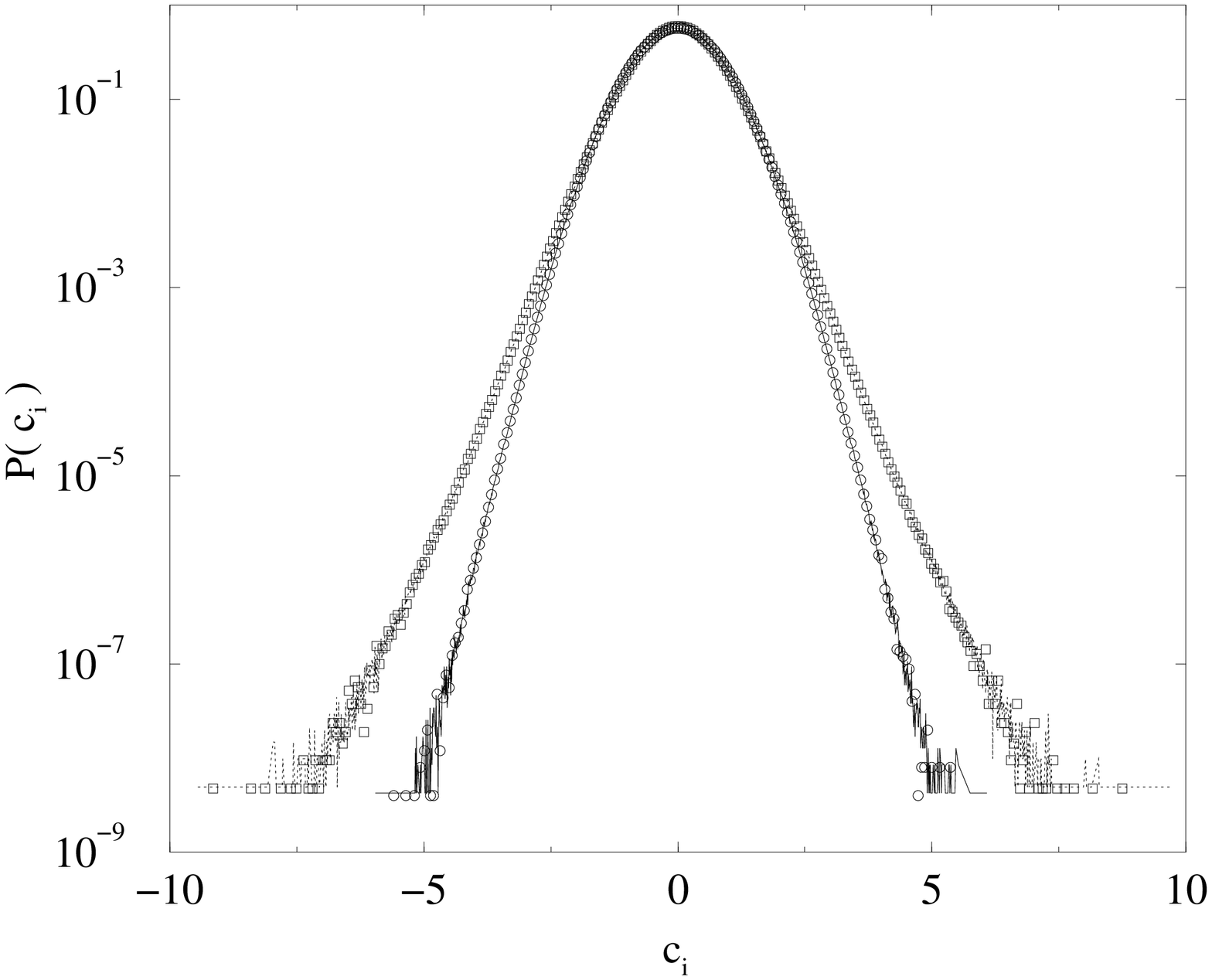,width=6cm,angle=-90}
  \vspace{0.5cm}}
\caption{Velocity distribution functions for DSMC simulations of the
random $\alpha$ model, for two sets of two distributions having the
same values of $\overline{\alpha}$ and $\overline{\alpha^4}$:
Trimodal $\rho(\alpha)=\frac{b}{2} 
(\delta(\alpha - \sqrt{1+\gamma}) +
\delta(\alpha - \sqrt{1-\gamma})) + (1-b)\delta(\alpha - 1)$ with
$(b \approx 0.546248, \gamma \approx 0.390584)$ (solid line) versus
a flat distribution for $\alpha^2 \in [0.5;1.5]$ (circles), and
another trimodal distribution with
$(b \approx 0.47779,\gamma \approx 0.835254)$ (dotted line) versus
another flat distribution for $\alpha^2 \in [0:2]$ (squares). The
agreement over $8$ orders of magnitude shows that the $a_2$ approximation
yields reliable predictions.
}
\label{fig:trimodal_flat}
\end{figure}

The theoretical (from eq (\ref{eq:a2}))
and measured (from the fourth moment of the velocity distribution)
values of $a_2$ are given in Table 1. For small $a_2$ the measured 
and theoretical values
agree perfectly, while for larger $a_2$ the agreement is worse, probably
due to the neglect of quadratic $a_2^2$ and higher order terms   
in the derivation of the theoretical formula. In fact the disagreement 
is at most of $10 \%$ when $|a_2| >0.1$, 
comparably to the results for constant $\alpha$.
As predicted, the same value of the fourth cumulant $a_2$ is associated with
distributions having the same first, second 
and fourth moments [see Eq. (\ref{eq:a2bis})]. 
Moreover, Fig. \ref{fig:trimodal_flat} shows strong numerical evidence
that, {\em in this case},
the whole distributions of velocities are indistinguishable, i.e. not only
$a_2$ but the complete statistics (including large velocities)
depend only on the first moments of $\rho(\alpha)$. Of course, for
other $\rho(\alpha)$, this property may not be true anymore.

\begin{figure}
\centerline{
       \psfig{figure=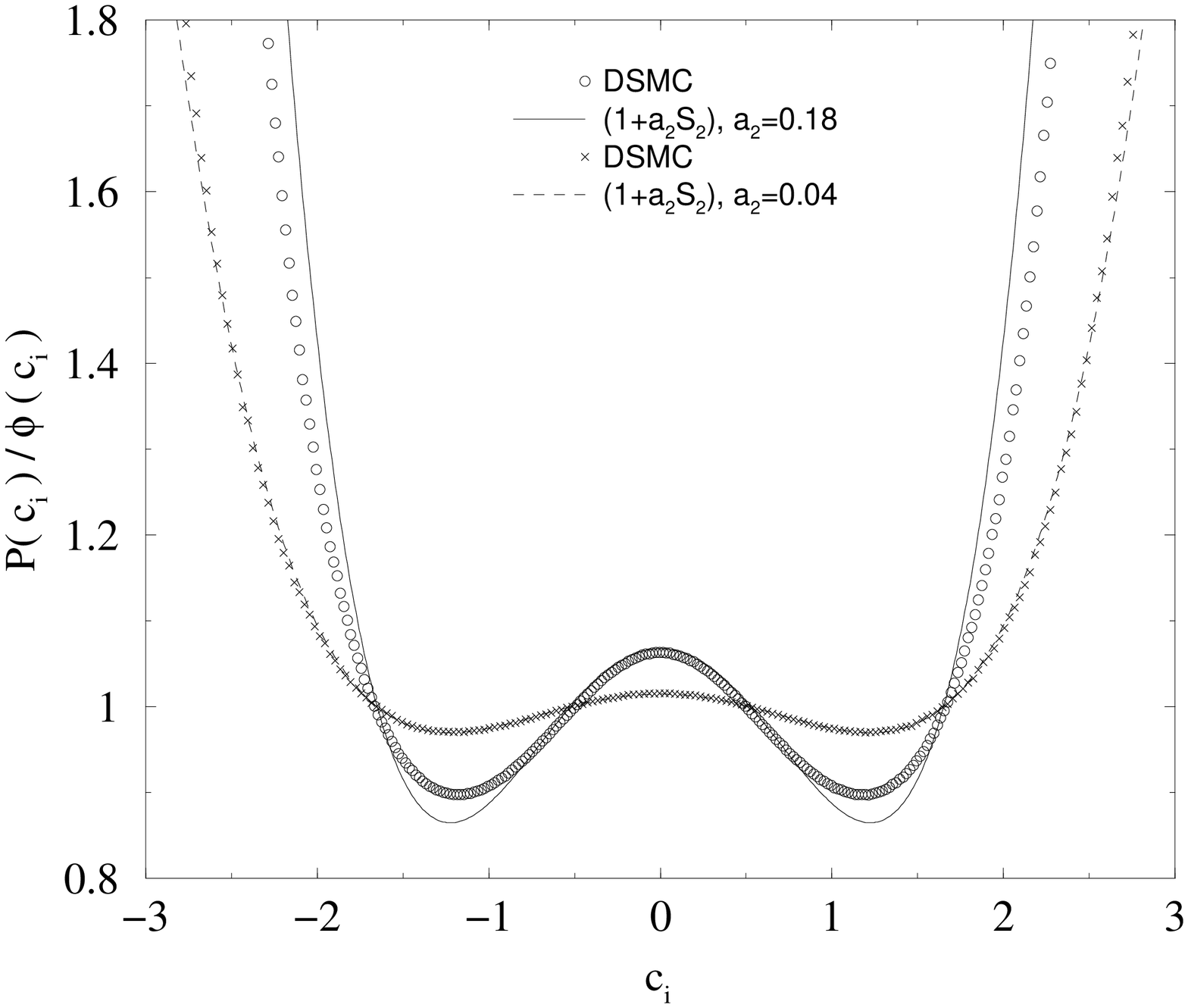,width=6cm,angle=-90}
  \vspace{0.5cm}}
\caption{Comparison of the ratios $\tilde{f}/\Phi$, measured
in DSMC simulations (symbols), versus the theoretical predictions
of the Sonine approximation $1+a_2 S_2$ (lines), for two flat distributions
of $\alpha^2$: $\alpha^2 \in [0,2]$ (circles) yields a relatively large 
fourth cumulant ($a_2 \approx 0.18$)
and the agreement is satisfying at small velocities only,
while the measured and theoretical curves are indistinguishable for
$\alpha^2 \in [0.5;1.5]$ (crosses), for which $a_2 \approx 0.04$ is small.
}
\label{fig:sonine}
\end{figure}

In fig. \ref{fig:sonine}, we show 
the comparison between the normalized velocity distribution 
for two distributions of restitution coefficients, 
together with the Sonine prediction $1+a_2 S_2$.
In the case of small $a_2$, the agreement is striking: the
Sonine expansion is supposed to be valid for small values
of $c$, while we see in Fig. \ref{fig:sonine} an agreement
up to relatively large $c$. For larger
$a_2$, the global shape corresponds to the prediction, but the quantitative
agreement is lost. This is quite expected since, when $a_2$ is
not very small, higher order terms in the Sonine expansion become
relevant.

\vskip -1cm

\begin{figure}
\centerline{
       \psfig{figure=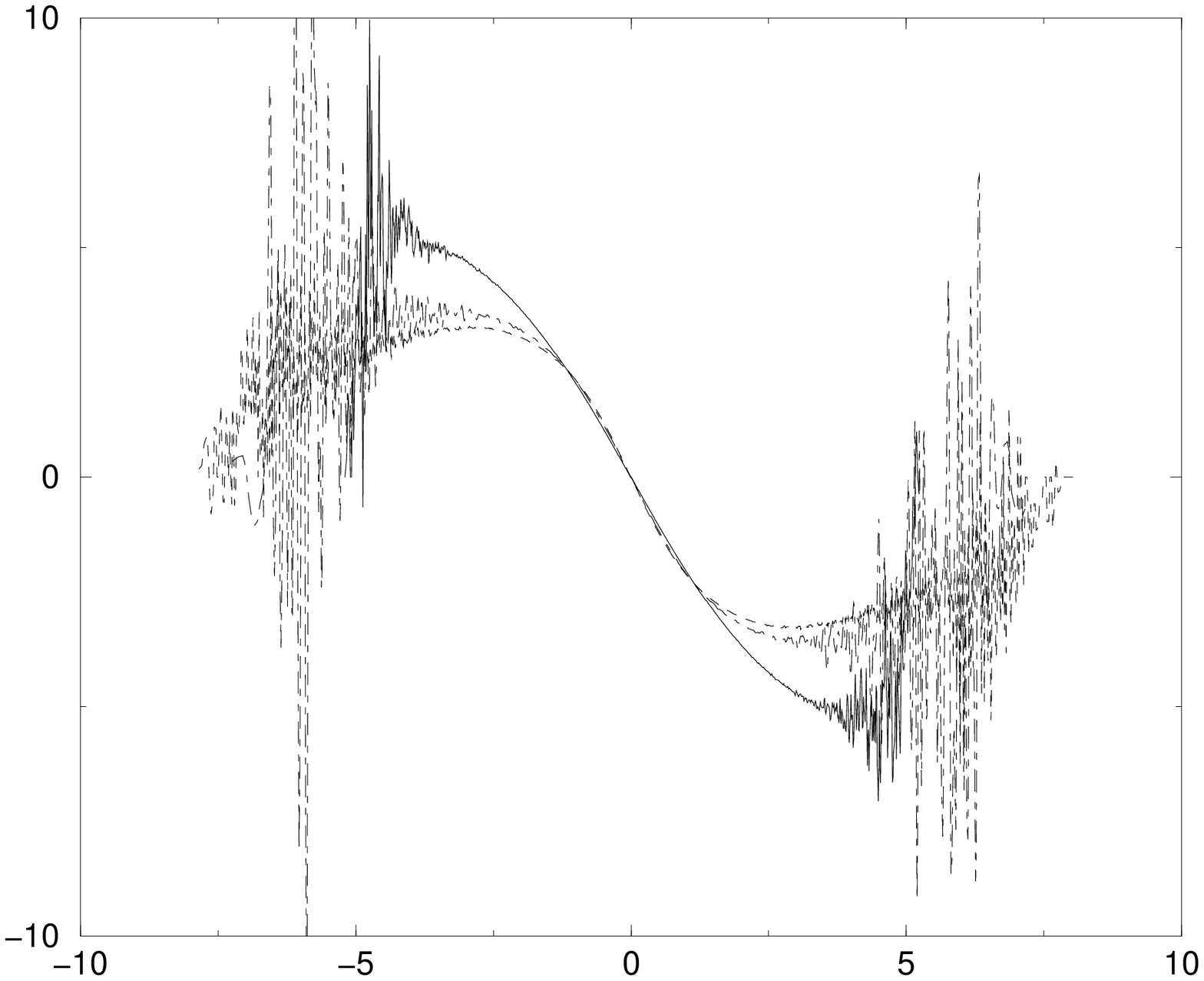,width=6cm,angle=-90}
  \vspace{0.5cm}}
\caption{Derivative $d\ln \tilde{f}/dc$ versus $c$ for various $\rho(\alpha)$:
from top to bottom on the right, flat $\tilde{\rho}(\alpha^2)$ 
with $\alpha^2 \in [0;2]$,
bimodal distribution $\tilde{\rho}(\alpha^2)=\frac{1}{2}
[\delta(\alpha^2-1/2)+\delta(\alpha^2-3/2)]$,
and flat $\tilde{\rho}(\alpha^2)$ in $[0.5,1.5]$.
}
\label{fig:derivlog}
\end{figure}

We now turn to the study of the large velocity tails. 
A first indication is given by the plot of the derivative of
$\ln \tilde{f}(c)$, which is linear for a Gaussian, and constant for an
exponential law. An $\exp(-A c^B)$ tail on the other hand leads
to a $c^{B-1}$ behaviour.
We see in Fig. \ref{fig:derivlog} that the non gaussianity
is indeed revealed by this criterion, 
but that the numerical noise hinders any clear
conclusion on the value of $B$. Since our velocity statistics  are smooth
over $8$ orders of magnitude, this approach is unable to determine
the values of $B$, even if $f(v)$ behaves asymptotically
as a stretched exponential.

\begin{figure}
\centerline{
       \psfig{figure=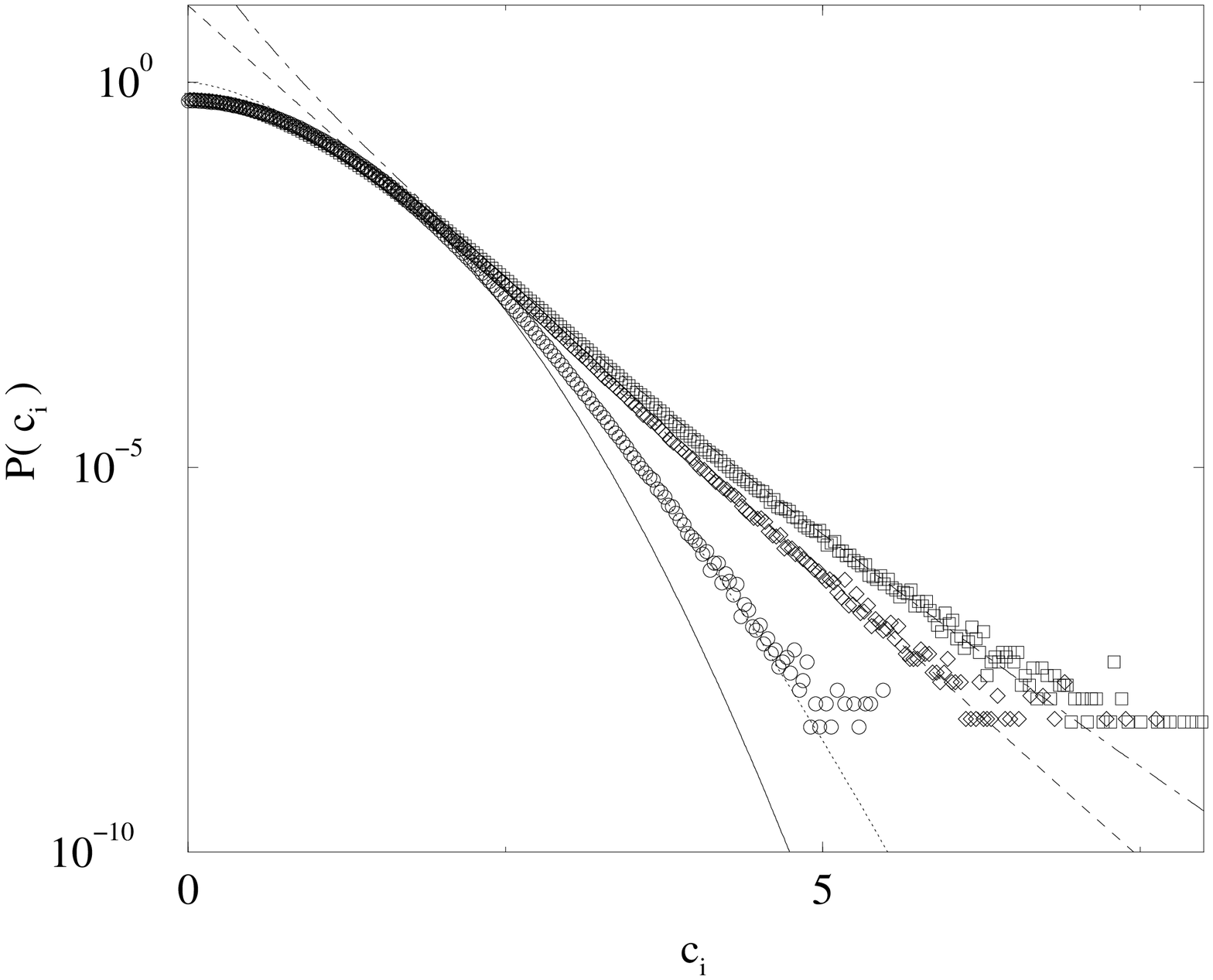,width=6cm,angle=-90}
  \vspace{0.5cm}}
\caption{Velocity distributions for the same $\rho(\alpha)$
as in Fig. \ref{fig:derivlog}
(circles: flat $\tilde{\rho}(\alpha^2)$ in $[0.5,1.5]$;
diamonds: bimodal $\alpha^2=1/2$ or $3/2$; squares:
flat $\tilde{\rho}(\alpha^2)$ in $[0;2]$), 
obtained from DSMC simulations, together
with the Maxwellian $\Phi$ (solid line) and fits to 
$K \exp(-A c^B)$ 
(dotted line: $\exp(-1.5 c^{1.6})$; dashed line: $10 \exp(-3.4 c)$;
dot-dashed line: $100 \exp(-5 c^{0.8})$).
The fits are accurate over 6 orders of magnitude
and values of $B$ consistent with the experimental data
($B \approx 1.6$) can be obtained for a convenient choice of $\rho(\alpha)$.
}
\label{fig:fits}
\end{figure}

Figure \ref{fig:fits} on the other hand
shows three fits, for three distributions of restitution coefficients.
These fits are of the form $\exp(-A c^B)$, and we obtain 
a wide range of possible values for $B$
\footnote{As stated above, for the situations investigated,
the value of $B$ seems to depend
on the first moments of $\rho(\alpha)$ only.}:
from $0.8$ to $2$, with fits accurate over
6 orders in magnitude. In particular, a convenient choice
of $\rho(\alpha)$ is compatible with $B=1.6$, which has been found in some
experiments~\cite{Losert,Rouyer} (close to
$B=3/2$ obtained in~\cite{twan} for randomly driven IHS fluids).
The corresponding $\rho(\alpha)$ [trimodal 
$\rho(\alpha)=\frac{b}{2} 
(\delta(\alpha - \sqrt{1+\gamma}) +
\delta(\alpha - \sqrt{1-\gamma})) + (1-b)\delta(\alpha - 1)$ with
$(b \approx 0.546248, \gamma \approx 0.390584)$, or equivalently
flat $\alpha^2 \in [0.5,1.5]$], is not very broad and
does not imply the use of particularly
large values of $\alpha$. Other choices of $\rho(\alpha)$ can also lead to
an exponential tail ($B=1$), similarly to the case of the
homogeneous cooling state (constant $\alpha < 1$, with no heating),
or even to larger distributions with $B < 1$.

\section{One-dimensional case}

For completeness, we briefly report the results obtained for the
one-dimensional version of the random-$\alpha$
model; it should correspond to a projection in one dimension of a 
two-dimensional system, with energy injection along the
projection direction.

The study of the projected model defined in section III
(injection of energy by randomly drawing 
the $y$-component of the velocity after each collision) yields
similar results as those obtained in the case of a three-dimensional 
system projected in 2D.

\begin{figure}
\centerline{
       \psfig{figure=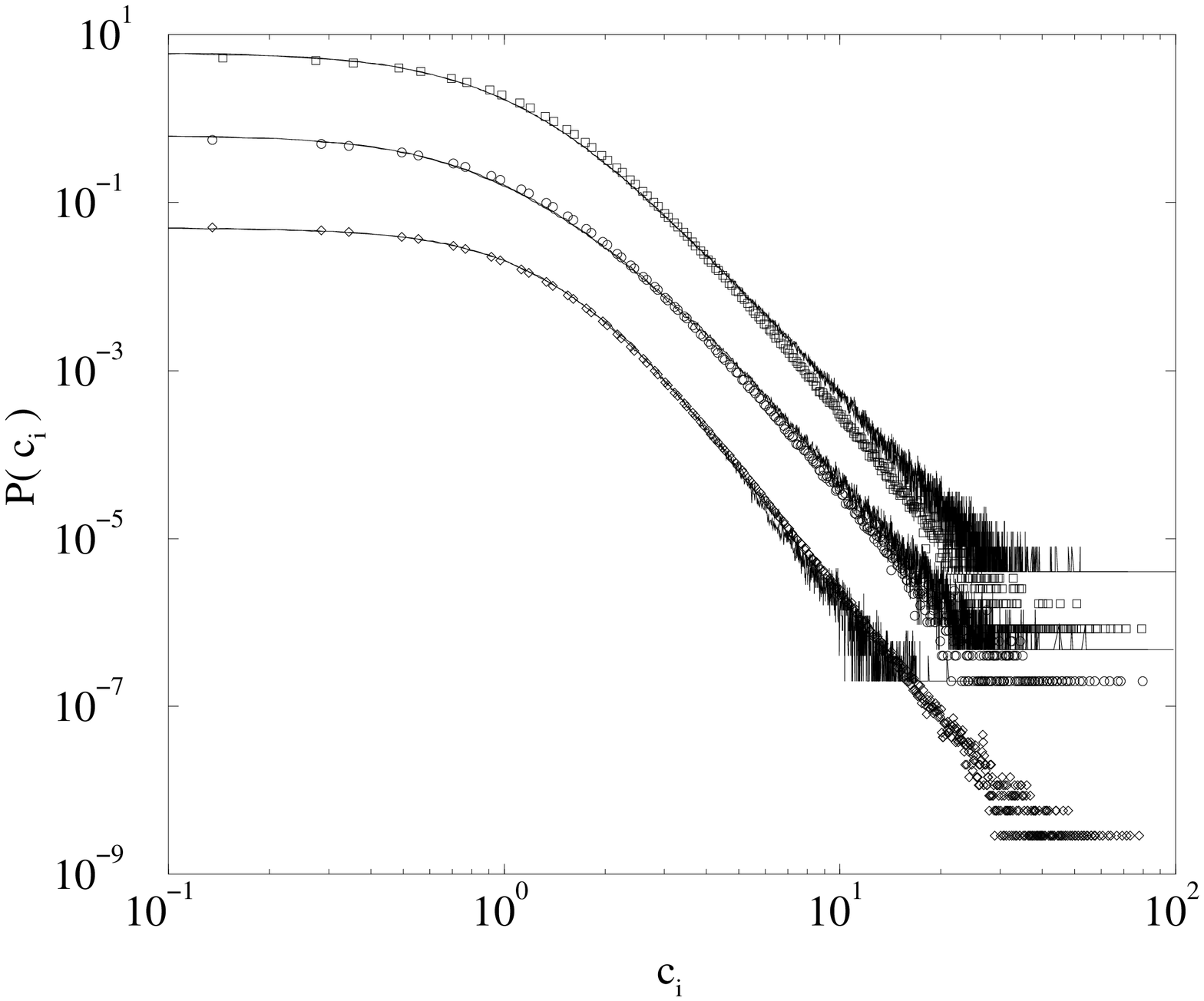,width=6cm,angle=-90}
  \vspace{0.5cm}}
\caption{Velocity distributions for the one-dimensional case,
for various $\rho(\alpha)$: flat $\rho$ (middle), flat 
$\tilde{\rho}$ ($\alpha^2 \in [0.5;1.5]$) (top), and bimodal
$\rho(\alpha)=\frac{1}{2}(\delta(\alpha - 1.04)
+\delta(\alpha - 0.958332)$ (bottom). Data for DSMC (symbols) and MD 
(lines) simulations are shown. Some distributions have 
been shifted for clarity.
}
\label{fig:1d}
\end{figure}

The random-$\alpha$ model however displays a certain number
of pathologies: for $d=1$ and $\overline{\alpha^2}=1$, we obtain
analytically $a_2=-16/15$ from Eq.(\ref{eq:a2bis}). 
This large value, independent of $\rho(\alpha)$ (i.e. non perturbative), 
indicates that the Sonine expansion is not valid.
In fact, numerical investigations (both MD and DSMC)
show that the velocity distributions $f(v)$ have
power law tails (see Fig. \ref{fig:1d}). 
This behaviour differs significantly from
a Gaussian and explains the failure of the Sonine approximation.

\section{Summary and Conclusions}

We have introduced the idea of a random restitution coefficient
in the IHS model to describe a vertically vibrated layer of 
granular material. Since energy is injected only along the
vertical axis, and transferred through collisions in the
perpendicular directions, our approach 
accounts for the fact (as shown in section III) 
that the projection in $2$ dimensions of a $3$-dimensional collision
can correspond to a gain in the two-dimensional energy, and therefore
to an effective restitution coefficient $\alpha$ larger than $1$, even if the
genuine $\alpha$ necessarily corresponds to a dissipative collision. 
The model is consequently studied
in 2 dimensions, with a probability distribution
$\rho(\alpha)$ for the restitution coefficient.
We have analyzed the velocity distributions, and in particular the deviation
from the Maxwellian, using the Sonine expansion technique. Following
\cite{twan} we obtained analytically the expression of the fourth 
cumulant $a_2$, 
which has been tested against Molecular Dynamics (MD) and Monte Carlo 
Direct Simulations (DSMC). 
It turns out that the theoretical predictions for $a_2$ 
are quite accurate, with a slight overestimation for $a_2$ that 
probably corresponds to the approximations made during the calculation
(nonlinear terms ${\cal O}(a_2^2)$ and higher order Sonine polynomials
neglected); in particular $a_2$  depends
only on the first moments of $\rho(\alpha)$; numerically, 
{\it the whole velocity distribution} has the same property with a very high
precision, at least for
the $\rho(\alpha)$ studied here (we do not exclude
that this behaviour could be violated by other kinds of
distributions). Moreover, the comparison 
between numerical data and the second order Sonine expansion shows a 
remarkable agreement for small values of $a_2$.
The high energy tails, studied with DSMC simulations, can be fitted
by functions of the form $\exp(-A c^B)$, with $B<2$ depending on
$\rho(\alpha)$. It would certainly be interesting to have theoretical
predictions concerning $B$; it seems for example that the high-energy tail
is always overpopulated with respect to the Maxwellian, while
a particular choice of heating can also yield an
under-population, as shown in~\cite{montanero}. Note that once a functional
form has been chosen for $\rho(\alpha)$, very different tails can be 
observed depending on the range of variation for $\alpha$
(compare the circles and squares in Fig. \ref{fig:fits}).
This feature might question the relevance of the exponent $B$ as 
an intrinsic quantity for granular gases in steady states. 

An interesting issue concerns the fact that no violation
of molecular chaos has been observed: MD and DSMC results are in remarkable
agreement (even with a packing fraction as high as 40\% in MD).
This is in contrast with the situation of free cooling~\cite{SLuding}
but also with MD results on heated inelastic hard spheres~\cite{Ignacio}. 
The microscopic precollisional velocity correlations driven by
the standard inelastic collision rule with a constant restitution
coefficient~\cite{Soto} seem significantly reduced within the present 
random $\alpha$ model. A thorough investigation of 
short scale velocity correlations would require the computation 
of various precollisional averages involving moments of the relative
velocities, and has not been performed. Our results however suggest
that the dynamical correlations inducing recollisions \cite{berne} 
and responsible for the violation of molecular chaos may not be a generic feature
of driven granular gases exhibiting a non equilibrium stationary state.

More refined models could introduce correlations
between the effective restitution coefficient and the relative
velocities of colliding pairs. This feature, neglected here, 
seems difficult to quantify from first principles, but might
affect the high energy tail or induce precollisional velocity 
correlations. It would however be very interesting to be able 
to link a realistic energy injection mechanism with a precise 
distribution of restitution coefficients.

Finally, a hydrodynamic study of the present random $\alpha$ model,
in which the conservation of the energy is valid on
average only, is left for future investigations.

\appendix

\section{Fourth cumulant of the velocity distribution}

In this appendix, we consider particular distributions $\rho(\alpha)$
(with $\overline{\alpha^2}=1$),
and give the corresponding formulae for $a_2$.

\begin{itemize}
\item Trimodal $\rho(\alpha)=\frac{b}{2} 
[\delta(\alpha - \sqrt{1+\gamma}) +
\delta(\alpha - \sqrt{1-\gamma})] + (1-b)\delta(\alpha - 1)$
\begin{equation}
a_2=\frac{16 b \gamma^2}
{8b(\sqrt{1+\gamma}+\sqrt{1-\gamma}) +32 -16b-15b\gamma^2}
\end{equation}

\item Bimodal $\rho(\alpha)=\frac{1}{2} 
(\delta(\alpha - \sqrt{1+\gamma}) +
\delta(\alpha - \sqrt{1-\gamma}))$ (particular case of the trimodal 
distribution, with $b=1$)
\begin{equation}
a_2=\frac{16\gamma^2}{16-15\gamma^2 +8(\sqrt{1+\gamma}+\sqrt{1-\gamma})}
\end{equation}
Example: for $\gamma=0.5$, $a_2 \approx 0.1444$

\item Flat distribution for $\alpha^2$, between $1-\gamma$ and $1+\gamma$
\begin{equation}
a_2=\frac{16\gamma^3}
{16\left((1+\gamma)^{3/2} - (1-\gamma)^{3/2} \right) +48\gamma -15\gamma^3}
\end{equation}
Example: for $\gamma=0.5$, $a_2 \approx 0.0436$; for
$\gamma=1$, $a_2 \approx 0.2045$.

\item Flat distribution for $\alpha$, between $\gamma$ and $1+\gamma$

$\overline{\alpha^2}=1$ imposes $\gamma=(\sqrt{11/3}-1)/2$, and yields
$a_2 \approx 0.1868$.
\end{itemize}

\section{Gaussian iff $\rho = \delta (\alpha -1 )$}
We show in this appendix that only the trivial elastic distribution
$\rho(\alpha)= \delta (\alpha -1 )$ can lead to Gaussian velocity
statistics.

Since we assume $\overline{\alpha^2}=1$, $\mu_2=0$ (see section III-A). 
Therefore, equation
(\ref{eq:ftilde}) reduces to
\begin{equation}
\int d\alpha \rho(\alpha) \tilde{I}(\tilde{f},\tilde{f}) =0 \ .
\label{eq:app1}
\end{equation}
Let us assume that $\tilde{f}(c)$ is Gaussian. The equations
for ${\mathbf v}_1^{**}$ and ${\mathbf v}_2^{**}$ lead to
\begin{equation}
\tilde{f}(c_1^{**})\tilde{f}(c_2^{**})=
\tilde{f}(c_1)\tilde{f}(c_2)\exp \left(-\frac{\alpha^{-2} -1}{2}
({\mathbf c}_{12} \cdot \bbox{\hat{\sigma}})^2 \right) \ .
\end{equation}
Equation (\ref{eq:app1}) is then recast, by carrying out the
integration over $\alpha$ for
the term $\tilde{f}(c_1)\tilde{f}(c_2)$, and
simplifying by $\tilde{f}(c_1)$, into
\begin{equation}
\gamma_d I_d\int d{\mathbf c}_2 c_{12} \tilde{f}(c_2) =
\int d\alpha \frac{\rho(\alpha)}{\alpha^2}
\int d{\mathbf c}_2 c_{12} \tilde{f}(c_2) J_d
\end{equation}
where
\begin{eqnarray}
\gamma_d I_d &=& \gamma_d 
\int_{-\pi/2}^{\pi/2} d\theta \sin^{d-2}\theta \cos\theta
\nonumber \\
J_d &=& 
\int' d\bbox{\hat{\sigma}}
\cos \theta 
\exp \left(-\frac{\alpha^{-2} -1}{2}
(c_{12} \cos \theta)^2 \right) \nonumber \\
&=& \gamma_d \int_{-\pi/2}^{\pi/2} d\theta \sin^{d-2}\theta
\cos \theta 
\exp \left(-\frac{\alpha^{-2} -1}{2}
(c_{12} \cos \theta)^2 \right) .
\end{eqnarray}
$\theta$ is the angle between $\bbox{\hat{\sigma}}$ and ${\mathbf c}_{12}$, 
and
$\gamma_d$ is a geometrical factor corresponding to the integration
over the remaining angles.
By expanding the exponential, we thus obtain the relation {\it valid for any 
${\mathbf c}_1$}:
\begin{eqnarray}
0 = I_d \int d{\mathbf c}_2 c_{12} \tilde{f}(c_2) -
\int d\alpha \frac{\rho(\alpha)}{\alpha^2} \times \nonumber \\
\int d{\mathbf c}_2 \tilde{f}(c_2)
\sum_{p=0}^\infty  \left(\frac{1-\alpha^{-2}}{2}\right)^p c_{12}^{2p+1}
\int_{-\pi/2}^{\pi/2} d\theta \sin^{d-2}\theta \cos^{2p+1}\theta
\end{eqnarray}
Since this is valid for any ${\mathbf c}_1$, each term of the expansion
in powers of $c_{12}$ must be zero. For $p=0$ we obtain
\begin{equation}
\overline{\alpha^{-2}}=1 \ ,
\end{equation}
and for $p \ge 1$:
\begin{equation}
\overline{\alpha^{-2} (1-\alpha^{-2})^p}=0.
\end{equation}
A straightforward recurrence yields $\overline{\alpha^{-2p}}=1$ for any
$p\ge 0$, and thus
\begin{equation}
\rho(\alpha)=\delta(\alpha-1)
\end{equation}

\begin{table}[t]
\begin{tabular}{lccccc}
 $\rho(\alpha)$ & $\overline{\alpha}$ & $\overline{\alpha^4}$ & 
theoretical $a_2$ & measured $a_2$ &\\
  \hline
bimodal 1  &   $0.966$ & $5/4$ & $0.144$ & $0.13 $&\\
bimodal 2 &$0.999$  &$1.00666$  & $3.3\ 10^{-3}$ &$3.3\ 10^{-3} $& \\
flat $\alpha$ &$0.957$  &$1.31111$  & $0.187$ & $0.162$& \\
flat $\alpha^2 \in [0;2]$ & $0.943$ & $4/3$ & $0.2$ & $0.178$ & \\
trimodal 1 & $0.943$ & $4/3$ & $0.2$ & $0.178$ & \\
flat $\alpha^2 \in [0.5;1.5]$ & $0.989$ & $1.0833$ & $4.4\ 10^{-2}$ &
$4.2\ 10^{-2}$ & \\
trimodal 2 & $0.989$ & $1.0833$ & $4.4\ 10^{-2}$ & $4.2\ 10^{-2}$ & \\
Gaussian, $s=0.1$ & $0.986$ & $1.09997$ & $5.3\ 10^{-2}$ & $0.051$ & \\
Gaussian, $s=0.2$ & $0.968$ & $1.199$ & $0.11$ & $0.102$ & \\
\end{tabular}
\caption{
bimodal 1:$\frac{1}{2}(\delta(\alpha - \sqrt{1/2})
+\delta(\alpha - \sqrt{3/2})$ ;
bimodal 2:$\frac{1}{2}(\delta(\alpha - 1.04)
+\delta(\alpha - 0.958332)$ ;
flat $\alpha$: $\alpha \in [0.457427;1.457427]$
trimodal: $\rho(\alpha)=\frac{b}{2} 
(\delta(\alpha - \sqrt{1+\gamma}) +
\delta(\alpha - \sqrt{1-\gamma})) + (1-b)\delta(\alpha - 1)$;
trimodal 1: $\gamma \approx 0.835254 , b\approx 0.47779$ ;
trimodal 2: $\gamma \approx 0.39058, b\approx 0.546248$;
Gaussian: $\tilde{rho}(\alpha^2) \propto 
\Theta(\alpha^2) \exp(-(\alpha^2 -1)^2/(2s^2))$
}
\label{table}
\end{table}

\end{document}